\documentclass[aps,twocolumn,pre]{revtex4-2}
\usepackage{amsthm}
\usepackage{ulem}
\usepackage{graphicx}
\usepackage{amsfonts}
\usepackage{amsmath}
\usepackage{amssymb}
\usepackage{bbold}
\usepackage{bm}
\usepackage{enumerate}
\usepackage{color}
\usepackage{comment}

\begin{document}
\title{Mesoscopic theory of flocking with alignment and antialignment copying}
\author{Chunming Zheng}
\email{cmzheng@ynu.edu.cn}
\affiliation{School of Physics and Astronomy, Yunnan University, Kunming, 650091, China}

\begin{abstract}
We study a stochastic model of collective orientation dynamics in which individuals update their orientation through pairwise aligning or anti-aligning copying interactions. We analyze both annealed dynamics, where interaction types are chosen probabilistically at each update, and quenched dynamics, where individuals are permanently assigned to aligning or anti-aligning subpopulations. Starting from the microscopic master equation on the circle, we derive a closed mesoscopic description via a Fourier-mode expansion and a systematic large-$N$ expansion, obtaining closed Fokker–Planck equations and effective stochastic differential equations for the polarization. We show that competing alignment and anti-alignment suppress long-range polar order in the thermodynamic limit in both cases, while finite systems display nontrivial fluctuation-induced structure controlled by the interaction composition. Our results, validated by Gillespie simulations, establish an analytically tractable framework for collective dynamics characterized by competing copying rules and intrinsic noise.
\end{abstract}
\maketitle

\section{Introduction}

Collective motion is one of the most striking emergent behaviors exhibited by biological systems, exemplified by the coordinated maneuvers of bird flocks and fish schools. The spontaneous formation of coherent group motion, often termed flocking or schooling, arises from simple local interactions among individuals yet displays complex global order \cite{vicsek1995novel,toner1995long,couzin2002collective,bertin2006boltzmann,buhl2006disorder,ballerini2008interaction,cavagna2010scale,solon2013revisiting,feinerman2018physics}. Understanding how microscopic interaction rules give rise to mesoscopic dynamics and ultimately macroscopic organization remains a central challenge in the physics of active and living matter.

While alignment-based models are often studied at the deterministic or hydrodynamic level \cite{vicsek1995novel,toner1995long,couzin2002collective,bertin2006boltzmann}, a complementary line of research formulates these interactions at the stochastic, individual level using master equations, thereby explicitly incorporating demographic noise and enabling systematic coarse-graining to mesoscopic dynamics \cite{biancalani2014noise,dyson2015onset}. Within this framework, intrinsic fluctuations arising from the discrete nature of individuals can qualitatively alter system behavior, driving states and transitions that are absent in deterministic mean-field descriptions. This perspective has motivated the derivation of mesoscopic descriptions, e.g., Fokker--Planck or Langevin equations, directly from microscopic interaction rules. Notably, the underlying interaction structure in many such models closely parallels the voter model, a paradigmatic stochastic process in which agents update their states via imitation or copying \cite{liggett1999stochastic,castellano2009statistical}. Building on this connection, Jhawar \textit{et al.}~\cite{jhawar2020noise} showed experimentally that finite-size fluctuations alone can induce spontaneous polarization even in the absence of deterministic bias, highlighting the constructive role of noise in emergent collective order. Following this framework \cite{biancalani2014noise,jhawar2020noise}, we use the term "flocking" to denote the emergence of collective orientational order via stochastic alignment dynamics. Our model focuses exclusively on orientational degrees of freedom without spatial motion or density variations, providing a minimal mesoscopic description of stochastic flocking.

Beyond alignment mechanisms, collective systems can also exhibit interactions that favor opposite orientations, resulting in effective antialignment of velocity directions. Such competing interactions have been explored in a broad range of active-matter models, including single-species systems with effective antialignment \cite{grossmann2014vortex,zheng2022noise,escaff2024self}, multispecies Vicsek-type models with competing alignment preferences \cite{chatterjee2023flocking,kursten2025emergent,lardet2026flocking}, and systems with nonreciprocal interactions, where asymmetric couplings give rise to novel collective phases, spontaneous chiral motion, and complex spatiotemporal dynamics \cite{mangeat2025emergent,chen2024emergent,Martin2025transition,kreienkamp2024nonreciprocal}. These studies highlight that competing orientational interactions can fundamentally enrich collective dynamics beyond the standard alignment paradigm.

By contrast, much less is known about the corresponding problem in stochastic reaction-like copying models, where interactions occur through probabilistic imitation events and demographic fluctuations play an essential dynamical role. In particular, although mesoscopic theories have been successfully developed for purely aligning stochastic copying dynamics \cite{jhawar2020noise}, an analytical mesoscopic description that incorporates both alignment and anti-alignment copying interactions has, to our knowledge, not yet been established.

In this paper, we introduce and analyze a minimal stochastic model of collective orientation dynamics driven by competing alignment and antialignment interactions, together with spontaneous random turning. Our objective is to understand how competing copying rules modify the effective mesoscopic drift and fluctuation structure governing collective orientation. Starting from the microscopic master equation, we derive closed mesoscopic Fokker-Planck and Langevin equations for the polarization order parameter. Our formulation can be viewed as a vector generalization of the voter model: whereas the standard voter model describes the imitation of discrete opinions \cite{liggett1999stochastic,castellano2009statistical}, our model extends these rules to continuous angular variables with circular symmetry, incorporating both alignment and ``anti-voter'' (antialignment) interactions.

We investigate two distinct realizations of the interaction dynamics: an annealed model, where interaction rules are dynamically resampled at each encounter, and a quenched model, where individuals are permanently assigned as aligners or anti-aligners. Physical systems with rapidly fluctuating interaction topologies, e.g., dynamically rewired active networks, fast behavioral mode-switching, or transient physical contact, are naturally described by the annealed model, which also serves as an analytically tractable benchmark for dynamically fluctuating interactions. Conversely, the quenched model incorporates permanent individual identity and static behavioral heterogeneity, reflecting physical and biological systems with fixed functional roles, such as conformist-contrarian populations \cite{hong2011kuramoto}, social antivoter interactions \cite{galam2004contrarian,castellano2009statistical}, and excitatory-inhibitory networks \cite{brunel2000dynamics}.

\section{Annealed alignment and antialignment copying}
\label{sec:annealed}
In the annealed realization, we consider a system of $N$ agents, where each agent $i \in \{1, \dots, N\}$ is characterized by a continuous orientation $\theta_i \in [-\pi, \pi)$. The state of the system evolves through two stochastic processes. First, spontaneous random turning occurs at a rate $a$, where an agent $i$ resets its orientation to a new value $\theta^*$ drawn from a uniform distribution $\mathcal{U}(-\pi, \pi)$. This process ensures rotational symmetry in the absence of interactions. Second, pairwise interactions occur at a total rate $c$. In an interaction event, a 'focal' agent $i$ and a 'partner' agent $j$ are selected; agent $i$ then adopts the orientation of $j$ with probability $p$ (alignment) or the opposite direction $(\theta_j + \pi) \pmod{2\pi}$ with probability $1-p$ (anti-alignment). These rules are summarized by the following reaction scheme:
\begin{equation}
\begin{cases}
\theta_i \xrightarrow{a} \theta^* \in \mathcal{U}(-\pi, \pi) &\text{(random turning)}\\
\theta_i + \theta_j \xrightarrow{cp} 2\theta_j &\text{(alignment copying)} \\
\theta_i + \theta_j \xrightarrow{c(1-p)} (\theta_j + \pi)+\theta_j  &\text{(antialignment copying)}.
\end{cases}
\end{equation}

Numerical simulations of these microscopic processes are implemented using the Gillespie stochastic simulation algorithm \cite{gillespie1977exact}, which provides an exact realization of the underlying master equation. To obtain a mesoscopic description, we introduce the empirical orientation density
\begin{equation}
\varphi(x)=\frac{1}{N}\sum_{i=1}^{N}\delta(x-\theta_i),
\end{equation}
which represents the fraction of individuals with orientation $x$. To analyze the collective dynamics, we decompose the empirical density into a Fourier series, $\varphi(x) = \frac{1}{2\pi} \sum_k \varphi_k e^{-ikx}$. The complex amplitudes $\varphi_k = \int_{-\pi}^{\pi} \varphi(x) e^{ikx} dx$ serve as the fundamental order parameters, where $\varphi_0 = 1$ ensures normalization and $\varphi_1$ corresponds to the global polar order vector $\mathbf{m}$.

The stochastic dynamics of the microscopic reactions induces a Markovian evolution for the probability functional $P(\varphi,t)$ of observing a density configuration $\varphi(x)$ at time $t$. Following standard reaction-process methods \cite{van1992stochastic}, the corresponding functional master equation can be written as
\begin{equation}
 \frac{d}{dt} P(\varphi, t) = N \int_{-\pi}^{\pi} dx \int_{-\pi}^{\pi} dy \, \mathcal{Q}(\varphi; x, y) P(\varphi, t), 
 \label{Eq:Master}
\end{equation}
where the transition operator $\mathcal{Q}$ encapsulates the local field dynamics,
\begin{equation}
\mathcal{Q}= (\Delta_x^+ \Delta_y^- - 1) \varphi(x) \left[ \frac{a}{2\pi} + c\gamma(\varphi, y) \right].
\end{equation}
Here, $\gamma(\varphi, y) = p \varphi(y) + (1-p) \varphi(y-\pi)$ accounts for the mixed alignment ($p$) and anti-alignment ($1-p$) copying interactions. The step operators $\Delta_x^\pm$ denote the addition or removal of a particle at orientation $x$. In the large $N$ limit, we expand it to the second order in $1/N$ \cite{minors2018noise},
\begin{equation}
\Delta_x^\pm \approx 1 \pm \frac{1}{N} \sum_k e^{-ikx} \partial_k + \frac{1}{2N^2} \sum_{k,\ell} e^{-i(k+\ell)x} \partial_k \partial_\ell,
\end{equation}
where $\partial_k \equiv \delta/\delta\varphi_k$. Substituting this expansion and the Fourier decomposition $\varphi(x) = \frac{1}{2\pi} \sum_k \varphi_k e^{-ikx}$ into Eq.~\eqref{Eq:Master}, the spatial integrals over $x$ and $y$ are evaluated. Due to the orthogonality of the Fourier modes, $\int_{-\pi}^{\pi} e^{i(k-m)x} dx = 2\pi \delta_{k,m}$, the continuous field evolution collapses into a multivariate Fokker-Planck equation for the joint distribution of modes $P(\{\varphi_k\}, t)$,
\begin{equation}
    \partial_t P = -\sum_k \partial_k (A_k P) + \sum_{k,\ell} \partial_k \partial_\ell (B_{k\ell} P).
\end{equation}
The drift and diffusion coefficients are derived as
\begin{equation}
\begin{aligned}
A_k &= -[a+c(1-\Gamma_k)]\varphi_k,\\
B_{k\ell} &= \frac{1}{2N}
\left(
a\varphi_{k+\ell}
+c\left\{
[1+(-1)^{k+\ell}]\varphi_{k+\ell}
-2\Gamma_\ell\varphi_k\varphi_\ell
\right\}
\right),
\end{aligned}
\label{Eq:anneal_drift_diff}
\end{equation}
where we define the coupling weight $\Gamma_k = p + (1-p)(-1)^k$. For the polar mode ($k=1$), the coupling weight simplifies to $\Gamma_1 = 2p-1$, which directly governs the onset of collective order as a function of the alignment probability $p$.

Note that the deterministic drift $A_k$ in Eq.~\eqref{Eq:anneal_drift_diff} is linear in $\varphi_k$, meaning that each Fourier mode relaxes independently at rate $\omega_k = a + c(1-\Gamma_k)$. In particular, higher-order modes ($|k| \ge 2$) decay rapidly without feeding back into the polar mode ($k=1$). In the diffusion matrix $B_{k\ell}$, coupling to higher modes enters only via state-dependent noise terms of order $\mathcal{O}(1/N)$, which vanish in the large-$N$ limit. Thus, setting higher modes to their zero mean values yields a closed mesoscopic Langevin description for the polarization order parameter.

\subsection{Polar dynamics and noise-induced ordering}
\begin{figure*}
\centering
\includegraphics[width=7in]{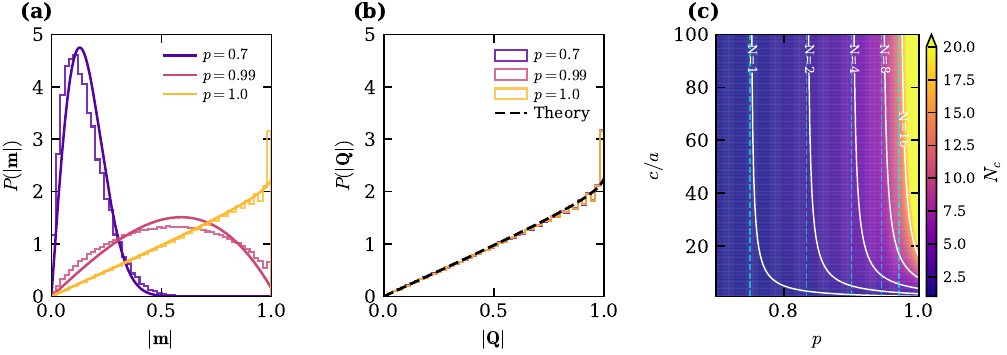}
\caption{
  Annealed dynamics results. 
  (a)~Probability density functions (PDFs) of the polar order parameter magnitude $|\mathbf{m}|$ for $N=50$, $a=0.1$, and $c=5.2$, illustrating the crossover across different alignment probabilities $p \in \{0.7, 0.99, 1.0\}$. Microscopic Gillespie simulations (step histograms) show excellent agreement with the stationary theory (solid curves); matching colors denote identical values of $p$.
  (b)~PDFs of the nematic order parameter magnitude $|\mathbf{Q}|$, demonstrating that the nematic order distribution is invariant under variations of $p$ and collapses onto the universal theoretical master curve (dashed line). 
  (c)~Phase diagram of the characteristic crossover system size $N_c$ in the $(p, c/a)$ parameter plane. Dashed lines indicate the theoretical asymptotes $p_c = (2N_c+1)/(2N_c+2)$ in the strong-coupling limit, delineating the boundaries of noise-induced ordering.  Simulation parameters and statistical sampling procedures are described in Appendix~\ref{Appendix:Gillespie}.}
\label{Fig:order_annealed}
\end{figure*}
To characterize the collective behavior, we track the dynamics of the polar order parameter $|\mathbf{m}|^2 = \varphi_1 \varphi_{-1}$. Using the identity for the time evolution of the expectation value of an observable $F(\boldsymbol{\varphi})$:
\begin{equation}
\frac{d}{dt} \langle F \rangle = \left\langle -\sum_k A_k \partial_k F + \sum_{k,\ell} B_{k\ell} \partial_k \partial_\ell F \right\rangle,
\end{equation}
where $\boldsymbol{\varphi} = (\dots, \varphi_{-1}, \varphi_0, \varphi_1, \dots)^\intercal$ denotes the set of Fourier modes and $\partial_k = \partial/\partial \varphi_k$. The brackets $\langle \dots \rangle$ denote an ensemble average over the stochastic realizations of the process, defined with respect to the density $P(\boldsymbol{\varphi}, t)$. We set $F = |\mathbf{m}|^2$ and thus the relevant derivatives are:
\begin{equation}
\begin{aligned}
\partial_k |\mathbf{m}|^2 &= \varphi_1 \delta_{k,-1} + \varphi_{-1} \delta_{k,1}, \\
\partial_k \partial_{\ell} |\mathbf{m}|^2 &= \delta_{k,-1} \delta_{\ell,1} + \delta_{\ell,-1} \delta_{k,1}.
\end{aligned}
\end{equation}

The drift contribution is found by summing over $k$, which yields $\langle A_{-1} \varphi_1 + A_1 \varphi_{-1} \rangle$. Given that $A_1$ and $A_{-1}$ share the same relaxation coefficient $\omega_c = a + 2c(1-p)$, this results in the linear term $-2 \omega_c \langle |\mathbf{m}|^2 \rangle$. The stochastic contribution is determined by the second mixed symmetric derivatives, which select the $B_{1,-1}$ and $B_{-1,1}$ components of the diffusion matrix. Importantly, for the present copying model, the diffusion coefficient associated with the polar mode depends only on the conserved zeroth Fourier mode ($\varphi_0=1$) and the polar order parameter $|\mathbf{m}|^2$. Consequently, higher Fourier modes are completely absent from the resulting mesoscopic equation, yielding an approximate closed description. This contrasts with kinetic theories of Vicsek-type active matter, where nonlinear collision terms generate an infinite hierarchy of angular moments requiring approximate closure schemes. The resulting equation of motion for the polar order is given by

\begin{equation}
\frac{\mathrm{d}}{\mathrm{d}t} \langle |\mathbf{m}|^2 \rangle = -2 \omega_c \langle |\mathbf{m}|^2 \rangle + \frac{2}{N} \langle \mathcal{D}(|\mathbf{m}|^2) \rangle,
\label{Eq:anneal_moment}
\end{equation}
where $\mathcal{D}(|\mathbf{m}|^2)
= a + c[1+(1-2p)|\mathbf{m}|^2]$ represents the state-dependent diffusion strength.

The moment equation \eqref{Eq:anneal_moment} is consistent with the following Itô Langevin equation for the polar vector $\mathbf{m}$:
\begin{equation}
\frac{d\mathbf{m}}{dt} = -\omega_c \mathbf{m} + \sqrt{\frac{\mathcal{D}(|\mathbf{m}|^2)}{N}} \boldsymbol{\eta}(t),
\end{equation}
where $\boldsymbol{\eta}(t)$ is Gaussian white noise. The steady-state distribution $P(|\mathbf{m}|)$ is obtained via the corresponding stationary Fokker-Planck equation,
\begin{equation}
P(|\mathbf{m}|) \propto |\mathbf{m}|
\left\{a+c[1+(1-2p)|\mathbf{m}|^2]\right\}^{\beta},
\label{Eq:annl_pol_distr}
\end{equation}
with the scaling exponent
\begin{equation}
\beta = -\frac{N[a + 2c(1-p)]}{c(1-2p)} - 1.
\end{equation}
The theoretical prediction in Eq.~\eqref{Eq:annl_pol_distr} is compared with microscopic simulations in Fig.~\ref{Fig:order_annealed}(a). As $p$ increases, anti-alignment is progressively suppressed, allowing alignment to continuously shift $P(\vert{}\mathbf{m}\vert{})$ toward higher order. For the parameters in Fig.~\ref{Fig:order_annealed}, the exponent governing the stationary distribution satisfies $\beta=0$ near $p_c \approx 0.9996$, which lies extremely close to unity. Thus, $p=0.99$ cleanly resolves the intermediate crossover where $P(\vert{}\mathbf{m}\vert{})$ broadens before state-dependent multiplicative fluctuations lead to boundary accumulation near $\vert{}\mathbf{m}\vert{} \to 1$ at $p=1.0$. The excellent agreement between simulation and theory confirms the validity of the mesoscopic Langevin description.
\subsection{Nematic order and coupling invariance}
The robustness of the nematic symmetry can be understood by examining how the interaction rules affect higher-order Fourier modes. Although the polar mode ($k=1$) is sensitive to the sign of the interaction via $\Gamma_1 = 2p-1$, as defined in Eq.~\eqref{Eq:anneal_drift_diff}, the nematic mode corresponds to $k=2$. Substituting this into our general expression for the coupling weight, we obtain:

\begin{equation}
\Gamma_2 = p + (1-p)(-1)^2 = 1.
\end{equation}

This identity holds for all values of $p$, reflecting the fact that the second Fourier mode is invariant under the $\pi$ shift inherent to antialignment. As a result, the deterministic drift for the nematic vector $\mathbf{Q}$ reduces to a simple relaxation $A_2 = -a\mathbf{Q}$, which is driven solely by the random turning rate $a$ and is independent of the copying strength $c$.

The stochastic properties of the nematic mode are similarly decoupled from the alignment fraction. By evaluating the diffusion coefficient $B_{2,-2}$ with $\Gamma_{-2}=1$, we find the state-dependent noise strength $\mathcal{D}(|\mathbf{Q}|^2) = a + c(1 - |\mathbf{Q}|^2)$. The resulting Langevin equation for the nematic vector $\mathbf{Q}$ is

\begin{equation}
\frac{d\mathbf{Q}}{dt} = -a \mathbf{Q} + \sqrt{\frac{a + c(1 - |\mathbf{Q}|^2)}{N}} \boldsymbol{\eta}(t).
\label{Eq:annl_nema_sde}
\end{equation}

Physically, this indicates that the nematic sector is "blind" to the competition between alignment and antialignment. This leads to the stationary distribution for the nematic magnitude:

\begin{equation}
P(|\mathbf{Q}|) \propto |\mathbf{Q}| \left[ a + c(1 - |\mathbf{Q}|^2) \right]^{\frac{Na}{c} - 1}.
\label{Eq:annl_nema_distr}
\end{equation}
This prediction is tested against simulations in Fig.~\ref{Fig:order_annealed}(b). In contrast to the polar case, the distributions for $p=0.7$ and $p=1.0$ collapse onto the same curve, confirming that the nematic order is invariant under changes in the alignment probability. The agreement between theory and simulation demonstrates that the nematic dynamics is fully captured by Eq.~\eqref{Eq:annl_nema_sde}, and depends only on the interaction strength $a$ and $c$ rather than its decomposition into alignment and anti-alignment components.

\subsection{Characteristic crossover system size and ordering boundaries}
In the present system, the emergence of collective motion is a purely noise-induced phenomenon, as the deterministic mean-field dynamics always favors the disordered state. Rather than defining a thermodynamic critical point, we introduce a characteristic crossover system size $N_c$ through the condition that the scaling exponent $\beta$ in Eq.~\eqref{Eq:annl_pol_distr} vanishes. At this crossover point, the distribution exhibits a linear dependence on the order parameter magnitude, i.e. $P(|\mathbf{m}|) \propto |\mathbf{m}|$, representing a state governed purely by the available geometric phase-space volume. The condition $\beta=0$ therefore marks the change in the qualitative shape of the stationary distribution rather than a genuine phase transition. Specifically, for $N < N_c$, multiplicative noise is sufficiently strong to sustain a polar peak near $|\mathbf{m}| \approx 1$, whereas for $N > N_c$, the system becomes relaxation dominant and the peak retreats toward the disordered regime.

Figure~\ref{Fig:order_annealed}(c) maps $N_c$ across the parameter space of the alignment probability $p$ and the dimensionless coupling ratio $c/a$. The ordered phase is most robust as $p \to 1$, where the characteristic crossover size $N_c$ becomes large; in this limit, even relatively large populations can maintain alignment, provided the coupling $c/a$ is sufficiently high. Conversely, as $p \to 0.5$, $N_c \to 0$, indicating that the noise-induced ordering mechanism disappears when alignment and anti-alignment events occur with equal frequency. For $p<0.5$, anti-alignment dominates over alignment, and the stationary distribution no longer exhibits the noise-induced polar ordered phase for any finite system size.

While the noise-induced ordering observed here bears a conceptual resemblance to noise-induced transitions in one-dimensional (1D) systems \cite{biancalani2014noise}, the 2D nature of the polar order parameter introduces a radial Jacobian $|\mathbf{m}|$. Unlike 1D models where transition thresholds often mark boundary divergences, this geometric factor ensures that $P(|\mathbf{m}|=0)$ remains zero. Consequently, $N_c$ should be interpreted as a crossover scale separating two qualitatively different stationary distributions: one dominated by the geometric phase-space factor and the other exhibiting a noise-induced polar peak.

The dashed lines in Fig.~\ref{Fig:order_annealed}(c) represent the theoretical asymptotes $p_c(N_c)$ derived in the strong-coupling limit ($c/a \to \infty$). In this regime, the condition $\beta = 0$ simplifies to

\begin{equation}
\beta \approx -\frac{2N_c(1-p)}{1-2p} - 1 = 0,
\end{equation}

yielding the characteristic crossover alignment probability

\begin{equation}
p_c(N_c) = \frac{2N_c + 1}{2N_c + 2}.
\label{Eq:p_critical}
\end{equation}

Equation~\eqref{Eq:p_critical} defines the asymptotic crossover boundary in parameter space. Below $p_c$, the deterministic drift towards disorder becomes insurmountable for the multiplicative fluctuations regardless of the coupling strength, so that the stationary distribution no longer supports the noise-induced polar ordered phase.

\section{Quenched alignment and antialignment copying}
\label{sec:quenched}
While the annealed realization assumes that interaction rules are selected stochastically at each encounter, many collective systems are characterized by persistent individual identities. In this quenched realization, agents are assigned fixed behavioral strategies—alignment or anti-alignment—that do not fluctuate over time. This transition from transient interactions to static behavioral heterogeneity mirrors the architecture of excitatory-inhibitory (E-I) networks in neuroscience \cite{brunel2000dynamics} and conformist-contrarian voter dynamics \cite{hong2011kuramoto}. It is worth noting that while nonreciprocal active matter often considers asymmetric interspecies coupling with same-species cooperation \cite{fruchart2021non}, our anti-aligners act as intrinsic contrarians that antialign with all neighbors globally, yielding a continuous-variable stochastic analog of conformist-contrarian voter dynamics where asymmetric interactions between fixed subpopulations drive novel dynamical phases.

We formalize this quenched heterogeneity by partitioning the $N$ agents into two static subpopulations: aligners (subpopulation 1, fraction $p$) and antialigners (subpopulation 2, fraction $1-p$). We denote their respective angular densities as $\varphi_1(x,t)$ and $\varphi_2(x,t)$, which are subject to fixed normalization constraints,
\begin{equation}
\int_{-\pi}^{\pi} \varphi_1(x) dx = p, \qquad \int_{-\pi}^{\pi} \varphi_2(x) dx = 1-p.
\end{equation}

The total density of the system remains the sum of these two contributions, $\varphi(x,t) = \varphi_1(x,t) + \varphi_2(x,t)$. Here, the fraction $p$ no longer represents a transition probability, but rather the fixed fraction of the aligner subpopulation.

Agents in subpopulation 1 strictly perform alignment copying, while agents in subpopulation 2 strictly perform anti-alignment ($\pi$-shifted) copying. The microscopic stochastic dynamics are governed by the following reaction scheme:
\begin{equation}
\begin{cases}
\theta_i^{(s)} \xrightarrow{a} \theta^* \in \mathcal{U}(-\pi, \pi) & \text{(random turning)},\\
\theta_i^{(1)} + \theta_j^{(s)} \xrightarrow{c} 2\theta_j^{(s)} & \text{(aligner copying)}, \\
\theta_i^{(2)} + \theta_j^{(s)} \xrightarrow{c} (\theta_j^{(s)} + \pi)+\theta_j^{(s)} & \text{(anti-aligner copying)},
\end{cases}
\end{equation}
where $s \in \{1, 2\}$ labels the subpopulation. This configuration introduces a two-field dynamics where the total polarization is determined by the vector sum of subpopulation orders.

The probability functional $P[\varphi_1,\varphi_2,t]$
evolves according to
\begin{equation}
\frac{d}{dt}P
=
N\sum_{s=1}^2
\iint dx\,dy\,
\left(
\Delta_{s,x}^+\Delta_{s,y}^- - 1
\right)
W_s(x,y)\,P,
\end{equation}
where the subpopulation-dependent transition rates $W_s(x,y)$ characterize an agent of subpopulation $s$ changing its orientation from $x$ to $y$,
\begin{align}
W_1(x,y)
&=
\varphi_1(x)
\left[
\frac{a}{2\pi}
+ c\,\varphi(y)
\right],
\\
W_2(x,y)
&=
\varphi_2(x)
\left[
\frac{a}{2\pi}
+ c\,\varphi(y-\pi)
\right].
\end{align}

Here, the parameter $a$ represents the rate of spontaneous random turning, while $c$ governs the pairwise copying process. Note that while the initiator of the interaction is subpopulation specific, the source of the new orientation is the total density $\varphi(y) = \varphi_1(y) + \varphi_2(y)$, reflecting the fact that agents interact with the entire population regardless of their identity.

\subsection{Mesoscopic dynamics of frozen subpopulations}
To derive the coarse-grained dynamics, we project the subpopulation densities onto a Fourier basis. We define the $n$th complex Fourier mode for subpopulation $s$ as

\begin{equation}
\varphi_{s,n}(t) = \int_{-\pi}^{\pi} e^{-inx} \varphi_s(x,t) \, dx.
\end{equation}

The zero-order modes are conserved and determined by the quenched fractions: $\varphi_{1,0} = p$ and $\varphi_{2,0} = 1-p$. The primary physical observables are the subpopulation polar vectors, $\mathbf{m}_s \in \mathbb{R}^2$, whose components are related to the first complex mode $\varphi_{s,1}$ via

\begin{equation}
\mathbf{m}_s = \begin{pmatrix} \text{Re}(\varphi_{s,1}) \\ \text{Im}(\varphi_{s,1}) \end{pmatrix} = \int_{-\pi}^{\pi} \begin{pmatrix} \cos(x) \\ \sin(x) \end{pmatrix} \varphi_s(x,t) \, dx.
\end{equation}

The total polar field is the vector sum $\mathbf{m} = \mathbf{m}_1 + \mathbf{m}_2$.

Using the standard $1/N$ expansion of the step operators $\Delta_{s,x}^\pm$ to second order, we describe the evolution of the probability functional through its moments. For any observable $F(\{ \varphi_{s,k} \})$, the expectation value evolves as

\begin{equation}
\frac{d}{dt} \langle F \rangle = \left\langle -\sum_{s,k} A_k^{(s)} \partial_{s,k} F + \sum_{s,k,\ell} B_{k\ell}^{(ss)} \partial_{s,k} \partial_{s,\ell} F \right\rangle,
\end{equation}
where $\partial_{s,k} = \frac{\partial}{\partial \varphi_{s,k}}$. Since the identities of the agents are frozen, fluctuations within each subpopulation originate from distinct sets of individuals. Consequently we neglect cross-diffusion terms $B^{(12)}$, treating the two subpopulations as independent noise sources.

Following the same approach as the annealed case, we arrive at the drift and diffusion terms for aligners ($s=1$),
\begin{equation}
\begin{aligned}
A_k^{(1)} &= -(a+c)\varphi_{1,k} + cp\varphi_k, \\
B_{k\ell}^{(1)} &= \frac{1}{N} \left[ \frac{a}{2} \varphi_{1,k+\ell} + \frac{c}{2} (2\varphi_{1,k+\ell} - 2p \varphi_{1,k} \varphi_\ell) \right].
\end{aligned}
\end{equation}
Similarly, for anti-aligners ($s=2$),
\begin{equation}
\begin{aligned}
A_k^{(2)} &= -(a+c)\varphi_{2,k}
+c(1-p)(-1)^k\varphi_k,\\
B_{k\ell}^{(2)} &= \frac{1}{N}
\biggl(
\frac{a}{2}\varphi_{2,k+\ell}
+\frac{c}{2}\{
[1+(-1)^{k+\ell}]\varphi_{2,k+\ell}\\
&\quad
+2(1-p)(-1)^\ell\varphi_{2,k}\varphi_\ell\}
\biggr).
\end{aligned}
\end{equation}
Let us define the normalized variables as $\mathbf{M}_1 = \mathbf{m}_1/p, \quad \mathbf{M}_2 =\mathbf{m}_2/(1-p)$. The total field remains $\mathbf{m} = p\mathbf{M}_1 + (1-p)\mathbf{M}_2$. Note that while $M_1 = m_1/p$ and $M_2 = m_2/(1-p)$ denote per-capita order parameters, $m_1$ and $m_2$ scale as $O(p)$ and $O(1-p)$ near the boundaries, respectively. Thus, the system remains nonsingular throughout $p \in [0,1]$. At the endpoints $p=1$ ($p=0$), the system smoothly contracts to the homogeneous single-subpopulation theory of pure aligners (anti-aligners) with $m_2=0$ ($m_1=0$). 
The evolution of the normalized subpopulation fields is governed by the following coupled Itô Langevin equations:
\begin{equation}
\begin{aligned}
\frac{d\mathbf{M}_1}{dt} &= -(a+c) \mathbf{M}_1 + c\mathbf{m} + \sqrt{\frac{[ a + c(1 - \mathbf{M}_1 \cdot \mathbf{m}) ]}{N_1}} \boldsymbol{\eta}_1(t), \\
\frac{d\mathbf{M}_2}{dt} &= -(a+c) \mathbf{M}_2 - c\mathbf{m} + \sqrt{\frac{[ a + c(1 + \mathbf{M}_2 \cdot \mathbf{m}) ]}{N_2}} \boldsymbol{\eta}_2(t),
\end{aligned}
\label{Eq:Quench_Langevin}
\end{equation}
where $N_1 = Np$ and $N_2 = N(1-p)$. The stochastic terms $\boldsymbol{\eta}_s(t)=(\eta_{s,x},\eta_{s,y})^\intercal$ are independent two-dimensional Gaussian white noises with $\langle \eta_{s,i}(t)\eta_{s',j}(t')\rangle =\delta_{ss'}\delta_{ij}\delta(t-t')$, where $s, s' \in \{1, 2\}$ denote the subpopulation index and $i, j \in \{x, y\}$ denote the Cartesian components. This independence reflects the quenched nature of the system, where demographic fluctuations within each subpopulation arise from distinct sets of agents.

\begin{figure}
\centering
\includegraphics[width=3.4in]{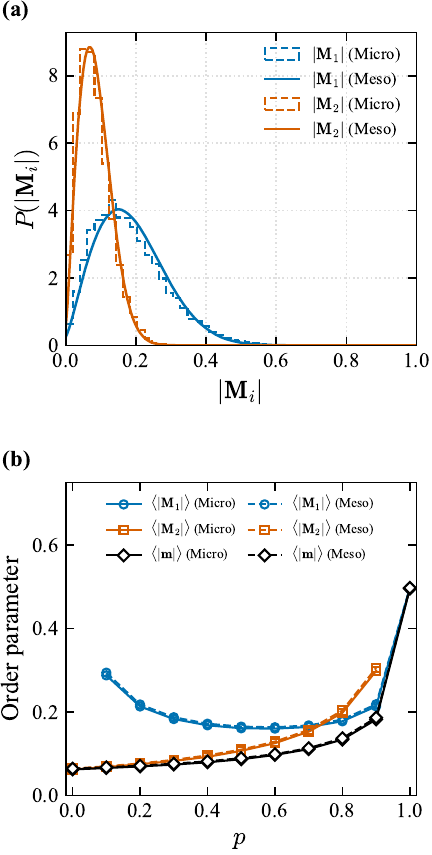}
\caption{Distribution of polarization magnitudes and stationary order parameters for the quenched alignment and antialignment model. (a) Probability density $P(|M_i|)$ of the aligner ($M_1$) and antialigner ($M_2$) population magnitudes. The stochastic microscopic agent-based results (dashed step-histograms) are compared against the mesoscopic Langevin dynamics (solid lines, obtained via kernel density estimation). Parameters are $N=100$, $p=0.3$, $a=0.5$ and $c=20$. (b) Stationary order parameters as a function of the aligner fraction $p$. Mean field magnitudes for the aligner population $\langle |M_1| \rangle$, the anti-aligner population $\langle |M_2| \rangle$, and the global system $\langle |m| \rangle$ are shown. Results from the microscopic model (solid lines, open markers) show excellent agreement with the mesoscopic approximation (dashed lines, open markers). Other parameters are consistent with (a). Error bars represent standard errors estimated from independent
realizations and are smaller than the symbol size and are therefore not visible on the scale of the figure.}
\label{Fig:Quenched}
\end{figure}

To validate the mesoscopic framework, we compare the stationary distributions of the sub-population polarization amplitudes, $|M_1|$ and $|M_2|$, obtained from the Langevin dynamics in Eq.~\eqref{Eq:Quench_Langevin} with direct microscopic agent-based simulations. As shown in Fig.~\ref{Fig:Quenched}(a), the two descriptions are in excellent agreement. Beyond this quantitative validation, the distributions reveal distinct statistical signatures that distinguish the two quenched subpopulations.

While both sub-populations are coupled to the same global field $\mathbf{m}$, the aligner field $|M_1|$ and anti-aligner field $|M_2|$ display significantly different peak positions and widths. This asymmetry originates from the quenched behavioral heterogeneity inherent to the quenched regime: because identities are frozen, each agent permanently retains its interaction type, perpetually contributing to either alignment or anti-alignment. Consequently, the two groups experience different effective fluctuation strengths driven by demographic noise, with the smaller sub-population typically exhibiting broader distributions due to finite-size effects ($D \propto 1/N_i$).

The dependence of the mean amplitudes on the fraction $p$ is summarized in Fig.~\ref{Fig:Quenched}(b). We observe an intriguing inverse trend: as $p$ increases, the aligner amplitude $\langle |M_1| \rangle$ slightly decreases while the anti-aligner amplitude $\langle |M_2| \rangle$ increases. This behavior is driven by the evolution of the global polarization field $\mathbf{m}$, which acts as the effective mean-field torque for both subpopulations.

As $p$ increases, the system becomes more globally ordered and the magnitude $|\mathbf{m}|$ grows. The anti-aligners respond to this strengthening global "anchor" by aligning more rigidly in the opposite direction. This stronger effective torque suppresses their angular fluctuations and increases their collective coherence, leading to the larger observed $\langle |M_2| \rangle$.

Conversely, the trend for the aligner sub-population is reversed due to a softening of this restoring torque. At low $p$, aligners are a minority fighting against a large population of anti-aligners; in this high-frustration environment, high internal cohesion is required to maintain a stable field. As $p$ increases and aligners begin to dominate ($\mathbf{m} \to \mathbf{M}_1$), the global field shifts from being an external constraint to a self-reinforcing reflection of the group’s own average. This relaxation of the effective restoring force allows for a slight increase in relative angular variance—despite the reduction in $1/N_1$ noise scaling—resulting in the observed reduction of $\langle |M_1| \rangle$.

The close agreement across the entire range of $p$ confirms that our stochastic field expansion accurately captures the subtle interplay between fixed heterogeneity and demographic noise in the quenched regime.

\subsection{Comparison with the annealed case}

In the limiting cases $p=1$ and $p=0$, the system effectively reduces to a homogeneous single population. Here, the diffusion coefficients read
\begin{equation}
B_{\mathrm{tot}} = \frac{a + c\bigl(1 \mp |\mathbf{m}|^2\bigr)}{N},
\end{equation}
where the minus and plus signs correspond to $p=1$ and $p=0$, respectively. In these limits, the quenched disorder disappears and the stochastic dynamics becomes identical to that of the annealed model, demonstrating the full consistency of the formulation.

For intermediate values $0 < p < 1$, the quenched system introduces an additional internal degree of freedom through the difference between the two subpopulation fields $\mathbf{M}_1$ and $\mathbf{M}_2$. However, for the total polar order parameter $\mathbf{m} = p \mathbf{M}_1 + (1-p) \mathbf{M}_2$, the resulting mesoscopic dynamics remains remarkably close to the annealed description.

At the deterministic level, the drift term for $\mathbf{m}$ is algebraically identical in both cases. Summing the weighted drifts from Eq.~\eqref{Eq:Quench_Langevin} yields 
\begin{equation}
 \mathbf{A}_{\text{tot}} = p \mathbf{A}_1 + (1-p) \mathbf{A}_2 = -a\mathbf{m} - 2c(1-p)\mathbf{m},   
\end{equation}
which is exactly the annealed drift. This indicates that both systems share the same mean-field evolution and the same deterministic relaxation structure.

Differences arise only at the level of fluctuations. In the quenched system, the total diffusion $B_{\text{q}}$ is the weighted sum of two independent noise sources,
\begin{equation}
\begin{aligned}
B_{\text{q}} &= \frac{p}{N} [a + c(1 - \mathbf{M}_1 \cdot \mathbf{m})] + \frac{1-p}{N} [a + c(1 + \mathbf{M}_2 \cdot \mathbf{m})]\\
&= \frac{a+c}{N} - \frac{c}{N} [p \mathbf{M}_1 \cdot \mathbf{m} - (1-p) \mathbf{M}_2 \cdot \mathbf{m}].
\end{aligned}
\end{equation}
By comparing this to the annealed diffusion $B_{\mathrm{a}} =\{a+c[1+(1-2p)|\mathbf{m}|^2]\}/N$, the discrepancy $\Delta B = B_{\text{q}} - B_{\text{a}}$ can be expressed in terms of the fluctuations $\delta \mathbf{M}_s$ of the subpopulations relative to their steady-state values. 

Each subpopulation obeys a mesoscopic Langevin equation with noise amplitude scaling as $N_s^{-1/2}$. As a consequence, fluctuations around the corresponding deterministic trajectories satisfy $\delta \mathbf M_s = O(N^{-1/2})$. Writing $\mathbf M_s = \bar{\mathbf M}_s + \delta \mathbf M_s$, where $\bar{\mathbf M}_s$ denotes the deterministic component, one can systematically expand the diffusion coefficient in powers of $\delta \mathbf M_s$.

\begin{figure}[t]
\centering
\includegraphics[width=3.4in]{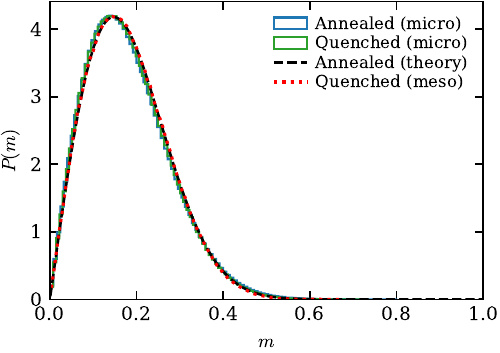}
\caption{Probability density $P(m)$ of the total order parameter $m=|\mathbf{m}|$ for annealed and quenched dynamics. Comparison of microscopic numerical simulations (histograms) with mesoscopic predictions for the quenched case (red dotted line) and the analytical annealed theory (black dashed line). Results are obtained under identical conditions with system size $N=50$, coupling parameters $a=0.5$ and $c=4$, and transition probability $p=0.8$.}
\label{Fig:order_compare}
\end{figure}

The correction to the annealed diffusion coefficient contains both linear and quadratic contributions in $\delta \mathbf M_s$. The linear terms scale as $O(N^{-3/2})$ after accounting for the overall $1/N$ prefactor, and vanish upon averaging since $\langle \delta \mathbf M_s \rangle = 0$. The leading nonzero contribution therefore arises from quadratic terms of the form $|\delta \mathbf M_s|^2$, whose expectation scales as $O(N^{-1})$. This yields an overall correction of order $O(N^{-2})$ to the diffusion coefficient. Consequently, the quenched and annealed diffusion coefficients satisfy
\begin{equation}
B_{\mathrm{q}}(\mathbf{m}) = B_{\mathrm{a}}(\mathbf{m}) + O(N^{-2}),
\end{equation}
demonstrating that the stochastic dynamics of the total polar order coincide up to subleading finite-size corrections. Since the leading noise amplitude itself is of the order of $O(N^{-1})$, these corrections are parametrically smaller. Consequently, no pronounced experimental signature in $P(\vert{}\mathbf{m}\vert{})$ distinguishes the two regimes, demonstrating the robustness of the collective state against individual strategy persistence.

This theoretical result is confirmed numerically in Fig.~\ref{Fig:order_compare}, where the probability distributions $P(m)$ for both dynamics remain nearly indistinguishable even for a moderate system size of $N=50$. The excellent agreement between the microscopic histograms and the mesoscopic theory (both quenched and annealed) validates the perturbative argument that the frozen disorder only subtly modifies the noise statistics of the total polar mode.

In contrast, the nematic mode exhibits an exact equivalence for all $N$ and $p$. Microscopically, an anti-alignment interaction maps an agent's orientation angle $\theta \to \theta + \pi$. In Fourier space, the $k$th angular mode transforms via the phase factor $e^{ik\pi} = (-1)^k$. For all even harmonics ($k=2, 4, \dots$), $(-1)^k = +1$, leaving these modes invariant under antialignment. Because nematic alignment ($k=2$) is quadrupolar and head-tail symmetric, anti-aligned pairs contribute identically to the second Fourier mode as aligned pairs. Consequently, both the deterministic drift and the noise amplitude of the nematic order parameter become independent of the disorder fraction $p$. More generally, this symmetry implies that all even Fourier modes of the angular distribution are unaffected by the choice of quenched versus annealed interactions, whereas only the odd (polar) modes [where $(-1)^k = -1$] are sensitive to the frozen heterogeneity of the agents.

\section{Conclusion}
\label{sec:conclusion}
We have analyzed a minimal non-equilibrium model with competing alignment and antialignment interactions, deriving mesoscopic Langevin equations from the microscopic master equation via a systematic $1/N$ expansion. Comparing annealed and quenched implementations of heterogeneity, we find that their deterministic drift terms are identical, indicating that macroscopic mean-field dynamics is insensitive to whether interaction rules are dynamically sampled or statically assigned. Differences arise only at the level of fluctuations: quenched disorder introduces persistent correlations that modify finite-size noise. These corrections are subleading in $1/N$ and are numerically negligible for moderate system sizes, thereby validating the annealed approximation as an effective description for finite populations.

A central result is the exact equivalence of the nematic sector in the two formulations. Because the second Fourier mode is invariant under the $\pi$ rotation associated with antialignment, the nematic dynamics are independent of the nature of the underlying heterogeneity. This symmetry-based decoupling contrasts with standard flocking models such as the Vicsek model, where nematic order is typically slaved to the polar mode. The present framework therefore highlights nematic order as an independent observable. Since it can be extracted from the same trajectory data, the simultaneous measurement of polar and nematic order provides a sensitive diagnostic for detecting competing interactions that may not be apparent from polarization alone.

Finally, the emergence of a characteristic crossover system size $N_c$ provides a quantitative illustration of the interplay between stochasticity and collective dynamics, echoing the "more is different" perspective of Anderson \cite{anderson1972more}. In this model, finite-size fluctuations can sustain order that is suppressed in the large-$N$ limit by deterministic decay. Importantly, this behavior arises in a system with competing and heterogeneous interactions, and our analysis shows that while annealed and quenched disorder lead to identical mean-field dynamics, they differ in their fluctuation structure at finite $N$. This identifies a regime in which both population size and the nature of microscopic heterogeneity jointly determine the stability of collective order. Such size-dependent crossover phenomena have also been the subject of recent investigations \cite{calvert2025many}. More broadly, our results provide a controlled framework to distinguish annealed and quenched disorder in nonequilibrium dynamics and to quantify their impact on fluctuations in biological and physical collectives.

\section*{Acknowledgements}
We acknowledge funding from the Yunnan Fundamental Research Projects (Grant No. 202401AU070216).

\appendix

\section{Numerical details of Gillespie simulations}
\label{Appendix:Gillespie}

To ensure reproducibility, the microscopic agent dynamics are implemented using a Gillespie-type event-driven algorithm based on the transition rates in Secs.~II and III. At each step, a spontaneous turning event occurs with probability $a/(a+c)$ (assigning a random phase from $[0, 2\pi)$ to a uniformly chosen particle), while a pairwise interaction occurs with probability $c/(a+c)$ between two randomly chosen particles. In the annealed model, the focal particle adopts its partner's orientation with probability $p$ or the opposite orientation with probability $1-p$. In the quenched model, particle identities (aligners vs. anti-aligners) are permanently assigned at $t=0$ according to fraction $p$. All simulations start from uniform random orientations on $[0, 2\pi)$.

For the annealed simulations (Fig.~1), each realization runs for $5 \times 10^6$ updates, discarding the first $4 \times 10^5$ steps as burn-in. Order parameters $|\mathbf{m}|$ and $|\mathbf{Q}|$ are sampled every 20 steps post-burn-in across 50 independent realizations. Empirical probability densities are calculated using normalized histograms with 50 uniform bins on $[0, 1]$.

For the quenched simulations (Fig.~2), each realization runs for $5 \times 10^6$ updates, discarding the first $10^5$ steps as burn-in, with sub-population magnitudes $|M_1|$ and $|M_2|$ sampled every $2 \times 10^4$ steps across 100 independent realizations. The corresponding mesoscopic Langevin equations are integrated using the Euler--Maruyama method with $\Delta t = 10^{-3}$ over $10^6$ steps ($10^5$ burn-in), sampling every 100 steps. Mesoscopic probability densities are estimated via Gaussian kernel density estimation with bandwidth $0.15$.

\normalem
\bibliography{references}

@article{vicsek1995novel,
  title={Novel type of phase transition in a system of self-driven particles},
  author={Vicsek, Tam{\'a}s and Czir{\'o}k, Andr{\'a}s and Ben-Jacob, Eshel and Cohen, Inon and Shochet, Ofer},
  journal={Physical review letters},
  volume={75},
  number={6},
  pages={1226},
  year={1995},
  publisher={APS}
}

@article{toner1995long,
  title={Long-range order in a two-dimensional dynamical XY model: how birds fly together},
  author={Toner, John and Tu, Yuhai},
  journal={Physical review letters},
  volume={75},
  number={23},
  pages={4326},
  year={1995},
  publisher={APS}
}

@article{buhl2006disorder,
  title={From disorder to order in marching locusts},
  author={Buhl, Camille and Sumpter, David JT and Couzin, Iain D and Hale, Joe J and Despland, Emma and Miller, Edgar R and Simpson, Steve J},
  journal={Science},
  volume={312},
  number={5778},
  pages={1402--1406},
  year={2006},
  publisher={American Association for the Advancement of Science}
}

@article{bertin2006boltzmann,
  title={Boltzmann and hydrodynamic description for self-propelled particles},
  author={Bertin, Eric and Droz, Michel and Gr{\'e}goire, Guillaume},
  journal={Physical Review E—Statistical, Nonlinear, and Soft Matter Physics},
  volume={74},
  number={2},
  pages={022101},
  year={2006},
  publisher={APS}
}

@article{ballerini2008interaction,
  title={Interaction ruling animal collective behavior depends on topological rather than metric distance: Evidence from a field study},
  author={Ballerini, Michele and Cabibbo, Nicola and Candelier, Raphael and Cavagna, Andrea and Cisbani, Evaristo and Giardina, Irene and Lecomte, Vivien and Orlandi, Alberto and Parisi, Giorgio and Procaccini, Andrea and others},
  journal={Proceedings of the national academy of sciences},
  volume={105},
  number={4},
  pages={1232--1237},
  year={2008},
  publisher={National Academy of Sciences}
}

@article{cavagna2010scale,
  title={Scale-free correlations in starling flocks},
  author={Cavagna, Andrea and Cimarelli, Alessio and Giardina, Irene and Parisi, Giorgio and Santagati, Raffaele and Stefanini, Fabio and Viale, Massimiliano},
  journal={Proceedings of the National Academy of Sciences},
  volume={107},
  number={26},
  pages={11865--11870},
  year={2010},
  publisher={National Academy of Sciences}
}

@article{solon2013revisiting,
  title={Revisiting the flocking transition using active spins},
  author={Solon, Alexandre P and Tailleur, Julien},
  journal={Physical review letters},
  volume={111},
  number={7},
  pages={078101},
  year={2013},
  publisher={APS}
}

@article{feinerman2018physics,
  title={The physics of cooperative transport in groups of ants},
  author={Feinerman, Ofer and Pinkoviezky, Itai and Gelblum, Aviram and Fonio, Ehud and Gov, Nir S},
  journal={Nature Physics},
  volume={14},
  number={7},
  pages={683--693},
  year={2018},
  publisher={Nature Publishing Group UK London}
}

@article{couzin2002collective,
  title={Collective memory and spatial sorting in animal groups},
  author={Couzin, Iain D and Krause, Jens and James, Richard and Ruxton, Graeme D and Franks, Nigel R},
  journal={Journal of theoretical biology},
  volume={218},
  number={1},
  pages={1--11},
  year={2002},
  publisher={Elsevier}
}

@book{van1992stochastic,
  title={Stochastic Processes in Physics and Chemistry},
  author={Van Kampen, NG},
  volume={1},
  year={1992},
  publisher={Elsevier}
}

@article{gillespie1977exact,
  title={Exact stochastic simulation of coupled chemical reactions},
  author={Gillespie, Daniel T},
  journal={The journal of physical chemistry},
  volume={81},
  number={25},
  pages={2340--2361},
  year={1977},
  publisher={ACS Publications}
}

@article{biancalani2014noise,
  title={Noise-induced bistable states and their mean switching time in foraging colonies},
  author={Biancalani, Tommaso and Dyson, Louise and McKane, Alan J},
  journal={Physical review letters},
  volume={112},
  number={3},
  pages={038101},
  year={2014},
  publisher={APS}
}

@article{dyson2015onset,
  title={Onset of collective motion in locusts is captured by a minimal model},
  author={Dyson, Louise and Yates, Christian A and Buhl, Camille and McKane, Alan J},
  journal={Physical Review E},
  volume={92},
  number={5},
  pages={052708},
  year={2015},
  publisher={APS}
}

@article{jhawar2020noise,
  title={Noise-induced schooling of fish},
  author={Jhawar, Jitesh and Morris, Richard G and Amith-Kumar, UR and Danny Raj, M and Rogers, Tim and Rajendran, Harikrishnan and Guttal, Vishwesha},
  journal={Nature Physics},
  volume={16},
  number={4},
  pages={488--493},
  year={2020},
  publisher={Nature Publishing Group UK London}
}

@article{grossmann2014vortex,
  title={Vortex arrays and mesoscale turbulence of self-propelled particles},
  author={Gro{\ss}mann, Robert and Romanczuk, Pawel and B{\"a}r, Markus and Schimansky-Geier, Lutz},
  journal={Physical review letters},
  volume={113},
  number={25},
  pages={258104},
  year={2014},
  publisher={APS}
}

@article{zheng2022noise,
  title={Noise-induced swarming of active particles},
  author={Zheng, Chunming and T{\"o}njes, Ralf},
  journal={Physical Review E},
  volume={106},
  number={6},
  pages={064601},
  year={2022},
  publisher={APS}
}

@article{chatterjee2023flocking,
  title={Flocking of two unfriendly species: The two-species Vicsek model},
  author={Chatterjee, Swarnajit and Mangeat, Matthieu and Woo, Chul-Ung and Rieger, Heiko and Noh, Jae Dong},
  journal={Physical Review E},
  volume={107},
  number={2},
  pages={024607},
  year={2023},
  publisher={APS}
}

@article{escaff2024self,
  title={Self-organization of anti-aligning active particles: Waving pattern formation and chaos},
  author={Escaff, Daniel},
  journal={Physical Review E},
  volume={110},
  number={2},
  pages={024603},
  year={2024},
  publisher={APS}
}

@article{kursten2025emergent,
  title={Emergent flocking in mixtures of antialigning self-propelled particles},
  author={K{\"u}rsten, R{\"u}diger and Mihatsch, Jakob and Ihle, Thomas},
  journal={Physical Review E},
  volume={111},
  number={2},
  pages={L023402},
  year={2025},
  publisher={APS}
}

@article{lardet2026flocking,
  title={Flocking beyond one species: Novel phase coexistence in a generalized two-species Vicsek model},
  author={Lardet, Eloise and Chen, Letian and Bertrand, Thibault},
  journal={Physical Review Letters},
  volume={137},
  number={11},
  pages={118303},
  year={2026},
  publisher={APS}
}

@article{brunel2000dynamics,
  title={Dynamics of sparsely connected networks of excitatory and inhibitory spiking neurons},
  author={Brunel, Nicolas},
  journal={Journal of computational neuroscience},
  volume={8},
  number={3},
  pages={183--208},
  year={2000},
  publisher={Springer}
}

@article{fruchart2021non,
  title={Non-reciprocal phase transitions},
  author={Fruchart, Michel and Hanai, Ryo and Littlewood, Peter B and Vitelli, Vincenzo},
  journal={Nature},
  volume={592},
  number={7854},
  pages={363--369},
  year={2021},
  publisher={Nature Publishing Group UK London}
}

@article{minors2018noise,
  title={Noise-driven bias in the non-local voter model},
  author={Minors, Kevin and Rogers, Tim and Yates, Christian A},
  journal={Europhysics letters},
  volume={122},
  number={1},
  pages={10004},
  year={2018},
  publisher={EDP Sciences, IOP Publishing and Societ{\`a} Italiana di Fisica}
}

@book{liggett1999stochastic,
  title={Stochastic Interacting Systems: Contact, Voter and Exclusion Processes},
  author={Liggett, Thomas M.},
  volume={324},
  year={1999},
  publisher={Springer-Verlag Berlin},
}

@article{galam2004contrarian,
  title={Contrarian deterministic effects on opinion dynamics:“the hung elections scenario”},
  author={Galam, Serge},
  journal={Physica A: Statistical Mechanics and its Applications},
  volume={333},
  pages={453--460},
  year={2004},
  publisher={Elsevier}
}

@article{castellano2009statistical,
  title={Statistical physics of social dynamics},
  author={Castellano, Claudio and Fortunato, Santo and Loreto, Vittorio},
  journal={Reviews of modern physics},
  volume={81},
  number={2},
  pages={591--646},
  year={2009},
  publisher={APS}
}

@article{hong2011kuramoto,
  title={Kuramoto model of coupled oscillators with positive and negative coupling parameters: an example of conformist and contrarian oscillators},
  author={Hong, Hyunsuk and Strogatz, Steven H},
  journal={Physical Review Letters},
  volume={106},
  number={5},
  pages={054102},
  year={2011},
  publisher={APS}
}

@article{anderson1972more,
  title={More is different: broken symmetry and the nature of the hierarchical structure of science.},
  author={Anderson, Philip W},
  journal={Science},
  volume={177},
  number={4047},
  pages={393--396},
  year={1972},
  publisher={American Association for the Advancement of Science}
}

@article{calvert2025many,
  title={How many more is different?},
  author={Calvert, Jacob and Richa, Andr{\'e}a W and Randall, Dana},
  journal={arXiv preprint arXiv:2510.06011},
  year={2025}
}

@article{mangeat2025emergent,
  title={Emergent complex phases in a discrete flocking model with reciprocal and non-reciprocal interactions},
  author={Mangeat, Matthieu and Chatterjee, Swarnajit and Noh, Jae Dong and Rieger, Heiko},
  journal={Communications Physics},
  volume={8},
  number={1},
  pages={186},
  year={2025},
  publisher={Nature Publishing Group UK London}
}

@article{chen2024emergent,
  title={Emergent chirality and hyperuniformity in an active mixture with nonreciprocal interactions},
  author={Chen, Jianchao and Lei, Xiaokang and Xiang, Yalun and Duan, Mengyuan and Peng, Xingguang and Zhang, Hepeng},
  journal={Physical Review Letters},
  volume={132},
  number={11},
  pages={118301},
  year={2024},
  publisher={APS}
}

@article{Martin2025transition,
  title = {Transition to Collective Motion in Nonreciprocal Active Matter: Coarse Graining Agent-Based Models into Fluctuating Hydrodynamics},
  author = {Martin, David and Seara, Daniel and Avni, Yael and Fruchart, Michel and Vitelli, Vincenzo},
  journal = {Phys. Rev. X},
  volume = {15},
  issue = {4},
  pages = {041015},
  numpages = {43},
  year = {2025},
  month = {Oct},
  publisher = {American Physical Society},
}

@article{kreienkamp2024nonreciprocal,
  title={Nonreciprocal alignment induces asymmetric clustering in active mixtures},
  author={Kreienkamp, Kim L and Klapp, Sabine HL},
  journal={Physical Review Letters},
  volume={133},
  number={25},
  pages={258303},
  year={2024},
  publisher={APS}
}

\end{document}